\documentstyle[psfig,aps,prc,eqsecnum]{revtex}

\begin{document}
\draft
\title{Three body force in peripheral $N\alpha$ scattering}
\author{R. Higa and M. R. Robilotta}
\address{Nuclear Theory and Elementary Particle Phenomenology Group\\
Instituto de F\'{\i}sica, Universidade de S\~{a}o Paulo,\\
Caixa Postal 66318, 05315-970, S\~{a}o Paulo, SP, Brazil}
\date{\today
}
\maketitle

\begin{abstract}
The exchange of a single pion does not contribute to $N\alpha$ 
scattering, due to the isoscalar nature of the target. Therefore 
peripheral $N\alpha$ scattering may be used to probe existing 
models for the two-pion exchange component of the $NN$ potential. 
In a recent work we have considered this possibility using an 
effective $N\alpha$ interaction based on two-body forces only. 
In the present communication we discuss the role played by the 
two-pion exchange three-nucleon force, since its range is 
comparable to the two-pion exchange potential (TPEP).
\end{abstract}


As in \cite{nalpha}, we assume the $\alpha$ ground state to 
be a S wave with a gaussian structure:

\begin{equation}
\big\vert \alpha \big\rangle=\Biggl({4\alpha\over\pi}\Biggr)
^{9/4}\,\exp\biggl(-2\alpha\sum_{i=1}^3\mbox{\boldmath ${\rho}$}
_i^2\biggr)\big\vert \chi_{\alpha} \big\rangle,
\label{eq:i.1}
\end{equation}

\noindent where $\rho_i$ are Jacobi coordinates, $\alpha$ is a 
parameter extracted from~\cite{kanada} and 
$\big\vert \chi_{\alpha} \big\rangle$ is the spin-isospin $\alpha$ 
wave function. The effective $N\alpha$ potential is given by

\begin{eqnarray}
\bar V&=&\bar V_{(2)}+\bar V_{(3)}\nonumber\\
&=&\big\langle \alpha \big\vert\sum_{i=1}^4V_{oi}\big\vert 
\alpha \big\rangle+\big\langle \alpha \big\vert\sum_{j>i=1}^4
\biggl[V_{oij}+V_{ijo}+V_{joi}\biggr]\big\vert \alpha \big\rangle,
\label{eq:i.2}
\end{eqnarray}

\noindent where $\bar V_{(2)}$ and $\bar V_{(3)}$ are due to 
two and three body forces respectively. The former was considered 
in \cite{nalpha}, and here we concentrate on the latter.

The two-pion exchange three-nucleon force involves the emission and 
absorption of virtual pions by nucleons $i$ and $j$, with an 
intermediate scattering by nucleon $k$. We use here the three body 
force derived in ref.~\cite{threebf}, where the intermediate $\pi N$ 
scattering is described in the framework of chiral symmetry, 
and given by

\begin{eqnarray}
V_{ijk}(\mbox{\boldmath $r$}_{ki},\mbox{\boldmath $r$}_{jk})
&=&\Biggl\{\Biggl({C_s\over\mu^2}\Biggr)\Bigl(\mbox{\boldmath 
${\tau}$}^{(i)}\cdot\mbox{\boldmath ${\tau}$}^{(j)}\Bigr)\Bigl(
\mbox{\boldmath ${\sigma}$}^{(i)}\cdot\mbox{\boldmath ${\nabla}$}
_{ki}\Bigr)\Bigl(\mbox{\boldmath ${\sigma}$}^{(j)}\cdot\mbox{
\boldmath ${\nabla}$}_{jk}\Bigr)\nonumber\\
&+&\Biggl({C_p\over\mu^4}\Biggr)\Bigl(\mbox{\boldmath ${\tau}$}
^{(i)}\cdot\mbox{\boldmath ${\tau}$}^{(j)}\Bigr)\Bigl(\mbox{
\boldmath ${\sigma}$}^{(i)}\cdot\mbox{\boldmath ${\nabla}$}_{ki}
\Bigr)\Bigl(\mbox{\boldmath ${\sigma}$}^{(j)}\cdot\mbox{\boldmath 
${\nabla}$}_{jk}\Bigr)\Bigl(\mbox{\boldmath ${\nabla}$}_{ki}\cdot
\mbox{\boldmath ${\nabla}$}_{jk}\Bigr)\nonumber\\
&-&\Biggl({C^{\prime}_p\over\mu^4}\Biggr)\Bigl(\mbox{\boldmath 
${\tau}$}^{(i)}\times\mbox{\boldmath ${\tau}$}^{(j)}\cdot\mbox{
\boldmath ${\tau}$}^{(k)}\Bigr)\Bigl(\mbox{\boldmath ${\sigma}$}
^{(i)}\cdot\mbox{\boldmath ${\nabla}$}_{ki}\Bigr)\Bigl(\mbox{
\boldmath ${\sigma}$}^{(j)}\cdot\mbox{\boldmath ${\nabla}$}_{jk}
\Bigr)\Bigl(\mbox{\boldmath ${\sigma}$}^{(k)}\cdot\mbox{\boldmath ${\nabla}$}_{ki}\times\mbox{\boldmath ${\nabla}$}_{jk}\Bigr)
\Biggr\}\nonumber\\
&\times&U(r_{ki})U(r_{jk}).
\label{eq:i.3}
\end{eqnarray}

In this result the function $U(x)$ is written as

\begin{equation}
U(x)={e^{-\mu x}\over\mu x}-
{\Lambda\over\mu}\;{e^{-\Lambda x}\over\Lambda x}
-{1\over 2}{\mu\over\Lambda}
\biggl({\Lambda^2\over\mu^2}-1\biggr)e^{-\Lambda x}
\label{eq:i.4}
\end{equation}

\noindent and the strenght coefficients have the numerical values 
$C_s=0.9204$, $C_p=-2.0082$ and $C^{\prime}_p=-0.6722$.

Due to the isoscalar nature of the $\alpha$ particle, the exchange 
of a single pion is not allowed and only $V_{ijo}$, which represents 
the external exchange of two pions, contributes to $\bar V_{(3)}$. 
It can be shown that the matrix element $\big\langle \chi_{\alpha} 
\big\vert\sigma^{(i)}_a \sigma^{(j)}_b\tau^{(i)}_c\tau^{(j)}_d
\big\vert \chi_{\alpha} \big\rangle$ vanishes for $a\neq b$ or 
$c\neq d$, and is equal to -1/3 for $a=b$ and $c=d$. Therefore, 
the term in $\bar V_{(3)}$ proportional to $C_p^{\prime}$ drops 
out and the remainder may be expressed as

\begin{equation}
\bar V_{(3)}(x)=\sum_{j>i=1}^4
\biggl[C_sI_{11}-{C_p\over 3}(I_{00}+3I_{22}-I_{20})\biggr],
\label{eq:i.5}
\end{equation}

\noindent where

\begin{eqnarray}
I_{mn}&=&16\sqrt 2\biggl({2\alpha\over\pi}\biggr)^3
\int d\mbox{\boldmath $u$}\;d\mbox{\boldmath $v$}\;{U_m(u)
\over u^n}{U_m(v)\over v^n}(\mbox{\boldmath $u$}\cdot
\mbox{\boldmath $v$})^n\nonumber\\
&\times&\exp\Bigl\{-2\alpha\bigl[3(\mbox{\boldmath $x$}-
\mbox{\boldmath $u$})^2+3(\mbox{\boldmath $x$}-\mbox{
\boldmath $v$})^2+2(\mbox{\boldmath $x$}-\mbox{\boldmath $u$})
\cdot(\mbox{\boldmath $x$}-\mbox{\boldmath $v$})\bigr]\Bigr\},
\label{eq:i.6}
\end{eqnarray}

\begin{eqnarray}
U_0(x)&=&{e^{-\mu x}\over\mu x}-{\Lambda\over\mu}
\;{e^{-\Lambda x}\over\Lambda x}
-{1\over 2}{\Lambda\over\mu}
\biggl({\Lambda^2\over\mu^2}-1\biggr)e^{-\Lambda x},
\label{eq:i.7}\\&&\nonumber\\
U_1(x)&=&-{e^{-\mu x}\over\mu x}\biggl(1+{1\over\mu x}\biggr)
+{\Lambda^2\over\mu^2}\biggl(1+{1\over\Lambda x}\biggr)
{e^{-\Lambda x}\over\Lambda x}
+{1\over 2}\biggl({\Lambda^2\over\mu^2}-1\biggr)e^{-\Lambda x},
\label{eq:i.8}\\&&\nonumber\\
U_2(x)&=&{e^{-\mu x}\over\mu x}\biggl(1+{3\over\mu x}
+{3\over\mu^2x^2}\biggr)\nonumber\\
&-&{\Lambda^3\over\mu^3}\biggl(1+{3\over\Lambda x}+{3\over
\Lambda^2x^2}\biggr){e^{-\Lambda x}\over\Lambda x}
-{1\over 2}{\Lambda\over\mu}
\biggl({\Lambda^2\over\mu^2}-1\biggr)e^{-\Lambda x}
\Bigl(1+{1\over\Lambda x}\Bigr).
\label{eq:i.9}
\end{eqnarray}

Using

\begin{equation}
(\mbox{\boldmath $u$}\cdot\mbox{\boldmath $v$})^n\exp\bigl
[+4\alpha(4\mbox{\boldmath $x$}-\mbox{\boldmath $u$})\cdot\mbox{
\boldmath $v$}\bigr]=\biggl(-{1\over 4\alpha}{d\over d\xi}\biggr)
^n\bigg|_{\xi=1}\exp\bigl[+4\alpha(4\mbox{\boldmath $x$}-\xi\mbox{
\boldmath $u$})\cdot\mbox{\boldmath $v$}\bigr]
\label{eq:i.10}
\end{equation}

\noindent and a little more algebra, we get our final expression 
for the effective potential due to three-body interaction

\begin{eqnarray}
\bar V_{(3)}(x)&=&6\biggl({128\alpha^2\sqrt 2\over\pi}\biggr)
\int_0^{\infty}du\int_{-1}^1d(-\cos\theta)
\int_0^{\infty}dv\;\exp\bigl[-2\alpha (8x^2-8xu\cos\theta+3u^2
+3v^2)\bigr]\nonumber\\
&\times&\Biggl\{-{C_s\over 4\alpha}uv^2\Bigl[uU_1(u)\Bigr]
\Bigl[vU_1(v)\Bigr]\biggl[{1\over uv^3}\Bigl(A^{+\prime}(\xi)-A^{-\prime}(\xi)\Bigr)\biggr]_{\xi=1}
-{C_p\over 3}(uv)^2U_0(u)U_0(v)\nonumber\\
&\times&\biggl[{1\over v}\Bigl(A^+(\xi)-A^-(\xi)\Bigr)\biggr]
_{\xi=1}-{C_p\over (4\alpha)^2}
\Bigl[u^2U_2(u)\Bigr]\Bigl[v^2U_2(v)\Bigr]
\biggl[{1\over u^2v^3}\Bigl(A^{+\prime\prime}(\xi)-A^{-\prime\prime}
(\xi)\Bigr)\biggr]_{\xi=1}\nonumber\\
&+&{C_p\over 3}\bigl[u^2U_2(u)\bigr]\bigl[v^2U_2(v)\bigr]\biggl
[{1\over v}\Bigl(A^+(\xi)-A^-(\xi)\Bigr)\biggr]_{\xi=1}\Biggr\},
\label{eq:i.11}
\end{eqnarray}

\noindent where

\begin{equation}
A^{\pm}(\xi)={e^{\pm 4\alpha v\sqrt{16x^2+\xi^2u^2+8xu(-\cos
\theta)}}\over \sqrt{16x^2+\xi^2u^2+8xu(-\cos\theta)}}.
\label{eq:i.12}
\end{equation}

The effective three body $N\alpha$ potential $\bar V_{(3)}$ is shown 
in fig.~\ref{fig1}. It is repulsive, and we expect its inclusion 
to decrease the curves of the phase shifts in~\cite{nalpha}.

As the three body force employed in this work~\cite{threebf} and the 
chiral potential~\cite{chiral} are built inside the same theoretical 
framework, we compare their results in fig.~\ref{fig2}, that 
shows the ratio of $\bar V_{(3)}$ and the central component of the 
effective chiral potential $\bar V_{(2)ch}$~\cite{nalpha}. 
Asymptotically, this ratio has a smooth variation, since both 
potentials have the same exponential pattern, differing only 
by a polynomial in $r$.

In fig.~\ref{fig3} we display the $N\alpha$ phase shifts, as functions 
of the laboratory energy, for the waves $D_{5/2}$ and $H_{11/2}$. 
The dashed lines represent our previous results~\cite{nalpha}, and 
the continuous lines, the inclusion of $\bar V_{(3)}$. Figure~\ref{fig4} 
exhibits the change $\Delta\delta$ due to the latter component. 
This quantity depends more on the $NN$ potential for waves with 
low angular momentum $L$, and this dependence tends to disappear 
for large values of $L$. In order to estimate the size of the 
three-body effect, we have replaced $\bar V_{(3)}$ by $\epsilon\,
\bar V_{(2)ch}$. For $\epsilon\approx 0.0485$ the differences 
$\Delta\delta$ for waves with large $L$ are well reproduced~
\cite{thesis}, giving a rough idea of the size of this effect.

In this work we have shown that three body force in peripheral 
$N\alpha$ scattering gives rise to a small effect, about 5\% 
of the two body interaction. Therefore the general features of 
the curves obtained previously are preserved. This shows that 
the three body force does not spoil the ability of peripheral 
$N\alpha$ scattering to discriminate the mid range part of $NN$ 
potential.

\begin{center}
{\bf ACKNOWLEDGMENT}
\end{center}

R. H. was supported by Funda\c {c}\~{a}o de Amparo \`{a}
Pesquisa do Estado de S\~{a}o Paulo (FAPESP).

\newpage
%

\begin{figure}[tbp]
\caption{The effective three body $N\alpha$ potential 
$\bar V_{(3)}(r)$. }
\label{fig1}
\end{figure}
\begin{figure}[tbp]
\caption{Ratio of $\bar V_{(3)}$ and the central component of 
the effective chiral potential $\bar V_{(2)ch}$. }
\label{fig2}
\end{figure}
\begin{figure}[tbp]
\caption{$N\alpha$ phase shifts for the waves (a) $D_{5/2}$ 
and (b) $H_{11/2}$, as functions of the laboratory energy.}
\label{fig3}
\end{figure}
\begin{figure}[tbp]
\caption{Countribution of three body force to $N\alpha$ 
phase shifts, $\Delta\delta$, for the waves (a) $D_{5/2}$ 
and (b) $H_{11/2}$.}
\label{fig4}
\end{figure}

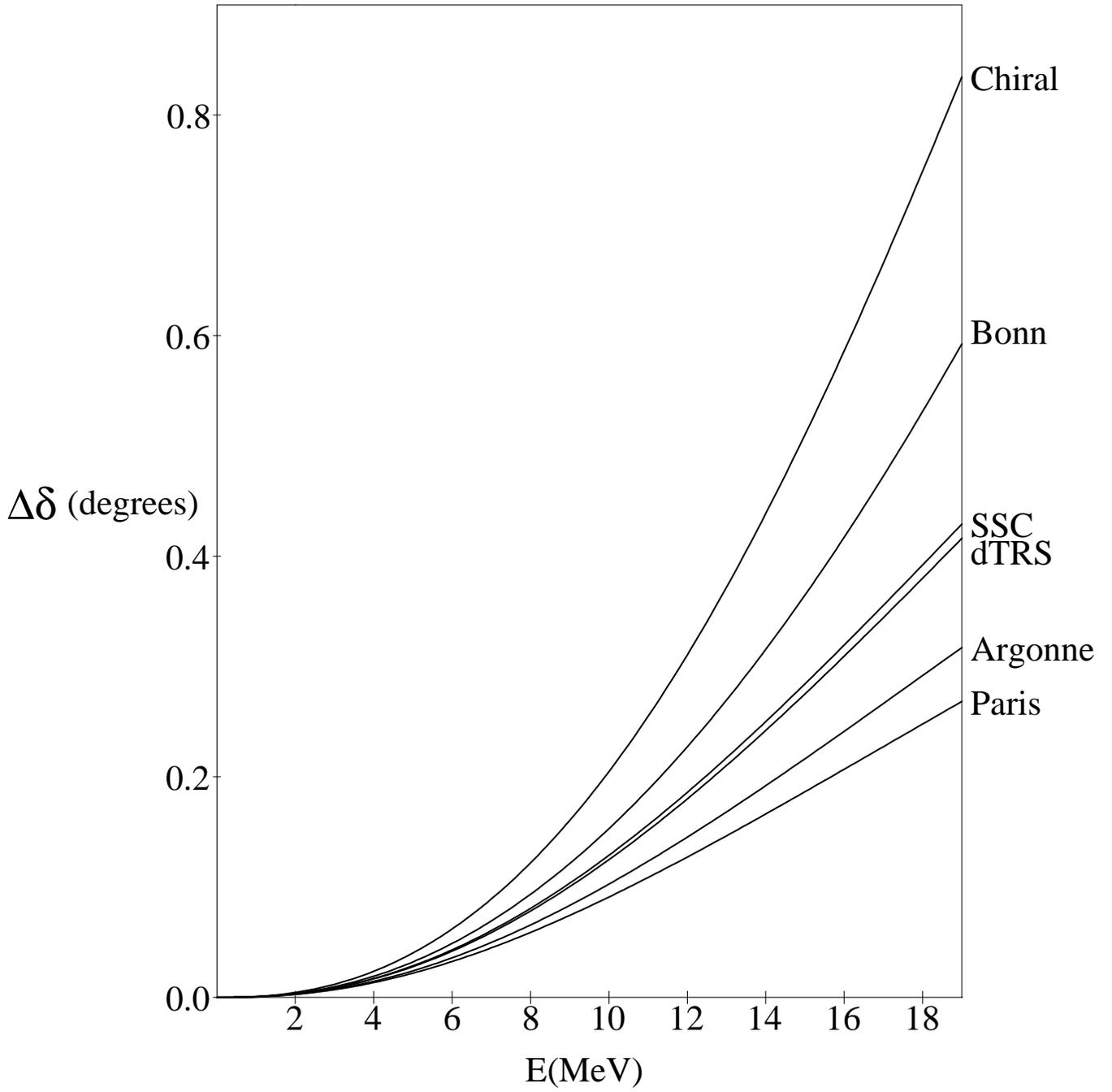
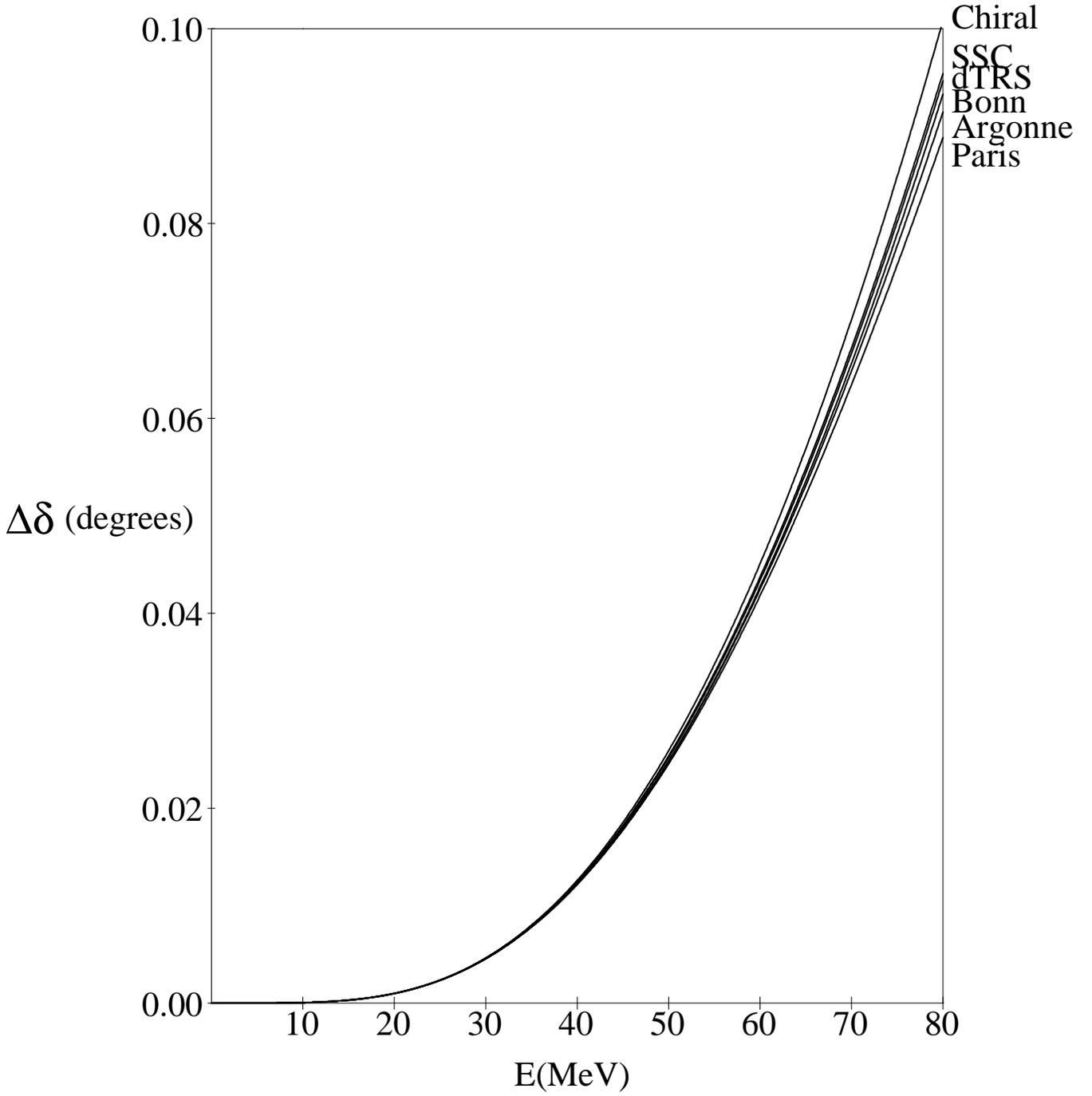
\end{document}